# On the Utility of Directional Information for Repositioning Errant Probes in Central Force Optimization

Richard A. Formato[1]

***Abstract.*** Central Force Optimization is a global search and optimization algorithm that searches a decision space by flying "probes" whose trajectories are deterministically computed using two equations of motion. Because it is possible for a probe to fly outside the domain of feasible solutions, a simple errant probe retrieval method has been used previously that does not include the directional information contained in a probe's acceleration vector. This note investigates the effect of adding directionality to the "repositioning factor" approach. As a general proposition, it appears that doing so does not improve convergence speed or accuracy. In fact, adding directionality to the original errant probe retrieval scheme appears to be highly inadvisable. Nevertheless, there may be alternative probe retrieval schemes that do benefit from directional information, and the results reported here may assist in or encourage their development.

*30 May 2010*
*Brewster, Massachusetts*

Ver. 2 (Fig. 1 improved for clarity; minor typos corrected.)
*6 June 2010*
*Brewster, Massachusetts*

[1]Registered Patent Attorney & Consulting Engineer
P.O. Box 1714, Harwich, MA 02645 USA
rf2@ieee.org

# On the Utility of Directional Information for Repositioning Errant Probes in Central Force Optimization

Richard A. Formato
Registered Patent Attorney & Consulting Engineer
P.O. Box 1714, Harwich, MA 02645 USA
rf2@ieee.org

## 1. Introduction

This note makes some observations about the utility of including directional information in the errant probe repositioning scheme used in Central Force Optimization [1-14]. CFO is a *deterministic* Nature-inspired global search and optimization metaheuristic that has been successfully applied to practical problems such as linear and circular array design [1,5], microstrip patch antenna design [15], and matching network optimization [1]. CFO also has performed well on recognized benchmark functions compared to other state-of-the-art algorithms [11-13].

CFO searches a decision space $\Omega$ by flying "probes" whose trajectories are based on a metaphor drawn from gravitational kinematics. Because CFO's probes may fly outside $\Omega$, a methodology is needed to deal with such errant probes. In previous implementations they were placed inside $\Omega$ using a simple deterministic scheme based on the "repositioning factor," $F_{rep}$, that does not include directional information. If any probe coordinate fell outside the decision space, then that coordinate was changed independently of the others, thereby losing the directional information contained in the acceleration that caused the probe to fly outside $\Omega$ in the first place.

It seems reasonable to speculate that retaining directionality would improve CFO's convergence by only truncating the errant probe's trajectory instead of changing its directional as well. This note takes a preliminary look at this question. It describes an errant probe methodology that combines the repositioning factor with acceleration directional information in order to investigate whether or not including directionality is beneficial. Perhaps somewhat surprisingly, not only does including directional information fail to improve convergence, in most cases it is an impediment.

## 2. The CFO Algorithm

CFO locates the maxima of an $N_d$-dimensional objective function $f(\vec{x})$ defined on a *decision space* of feasible solutions $\Omega : \{ \vec{x} \mid x_i^{\min} \leq x_i \leq x_i^{\max},\ 1 \leq i \leq N_d \}$, $x_i \in \Re$, where $\vec{x} = (x_1, x_2, ..., x_{N_d})$. $\Omega$ is bounded by $2N_d$ planes $P_{ik} : \{ \vec{x} \mid \vec{x} = (x_1,...,x_{i-1}, X_{ik}, x_{i+1},...,x_{N_d}) \}$ where $X_{ik} = \begin{cases} x_i^{\min}, & k=1 \\ x_i^{\max}, & k=2 \end{cases}$ (note that $i=1,...,N_d$; $k=1,2$ throughout).

CFO samples $\Omega$ by flying "probes" through it at a series of "time" steps (iterations). $f(\vec{x})$'s value, its *fitness*, is computed step-by-step at each probe's location which is specified by its position vector. At step $j-1$ probe $p$ is located at $\vec{R}_{j-1}^{p} = \sum_{i=1}^{N_d} x_i^{p,j-1} \hat{e}_i$ where $\hat{e}_i$ is the unit vector along the $i^{th}$ coordinate axis, $0 \leq j \leq N_t$ the iteration index, and $N_t$ the total number of steps (note that the first step is #0). $1 \leq p \leq N_p$ is the probe number and $N_p$ the total number of probes. Probe $p$ moves from the point $\vec{R}_{j-1}^{p}$ at step $j-1$ to $\vec{R}_j^p = \sum_{i=1}^{N_d} x_i^{p,j} \hat{e}_i$ at step $j$ under the influence of the (constant) acceleration $\vec{a}_{j-1}^p = \sum_{i=1}^{N_d} a_i^{p,j-1} \hat{e}_i$ produced by the CFO "masses" discovered by the probe distribution at step $j-1$.

Probe $p$'s motion in "CFO space" is computed from two *deterministic* "equations of motion" for the probe's trajectory and acceleration, respectively, as follows:

$$\vec{R}_j^p = \vec{R}_{j-1}^p + \vec{a}_{j-1}^p \tag{1}$$

$$\vec{a}_{j-1}^p = \sum_{\substack{n=1 \\ n \neq p}}^{N_p} U(M_{j-1}^n - M_{j-1}^p) \cdot (M_{j-1}^n - M_{j-1}^p) \times \frac{(\vec{R}_{j-1}^n - \vec{R}_{j-1}^p)}{\left\| \vec{R}_{j-1}^n - \vec{R}_{j-1}^p \right\|} \tag{2}.$$

$M_{j-1}^p = f(x_1^{p,j-1}, x_2^{p,j-1}, \ldots, x_{N_d}^{p,j-1})$ is the objective function's fitness at probe $p$'s location at time step $j-1$. Each of the other probes at that step (iteration) has associated with it fitness $M_{j-1}^n, n = 1, \ldots, p-1, p+1, \ldots, N_p$. $U(\cdot)$ is the Unit Step function defined as $U(z) = \begin{cases} 1, & z \geq 0 \\ 0, & otherwise \end{cases}$. Note that $\vec{a}_{j-1}^p = 0$ for $\vec{R}_{j-1}^n = \vec{R}_{j-1}^p, n \neq p$, because probe $n$ then has coalesced with probe $p$ and cannot exert any gravitational force on $p$. In this case the acceleration expression is indeterminate because $M_{j-1}^n = M_{j-1}^p$; and accordingly it is set to zero. Note that these equations have been simplified as described in [12,13].

### 3. The Errant Probe Problem

The acceleration computed from eq. (2) may large enough that trajectory eq. (1) flies probe $p$ outside the domain of feasible solutions, in which case it somehow must be returned to $\Omega$ or extinguished. While many return schemes are possible, the one that consistently has been used in previous CFO papers relies on either a fixed or variable *repositioning factor* $0 < F_{rep}^{min} \leq F_{rep} \leq 1$. Coordinate-by-coordinate, if the coordinate lies outside $\Omega$ then it is brought inside $\Omega$ as follows:

If $\vec{R}_j^p \cdot \hat{e}_i < x_i^{min}$ ∴ $\vec{R}_j^p \cdot \hat{e}_i = \max\{x_i^{min} + F_{rep}(\vec{R}_{j-1}^p \cdot \hat{e}_i - x_i^{min}), x_i^{min}\}$ (3a)

If $\vec{R}_j^p \cdot \hat{e}_i > x_i^{max}$ ∴ $\vec{R}_j^p \cdot \hat{e}_i = \min\{x_i^{max} - F_{rep}(x_i^{max} - \vec{R}_{j-1}^p \cdot \hat{e}_i), x_i^{max}\}$ (3b)

where the dot denotes vector scalar (inner) product. The out-of-bounds coordinate is set to a fraction $F_{rep}$ of the difference between its starting value and its boundary value. $F_{rep}$ starts at an arbitrary initial value $F_{rep}^{init}$ which then is incremented at each iteration by an arbitrary amount $\Delta F_{rep}$. If $F_{rep} > 1$ it is set to $F_{rep} = F_{rep}^{min}$, and the process continued. This procedure improves $\Omega$'s sampling by distributing errant probes throughout the decision space in an arbitrary but precise manner (see [10] for a more detailed discussion). At every step each probe's position is known exactly, so that every CFO run with the same setup parameters returns exactly the same results, thus preserving CFO's inherent determinism. What this scheme loses, however, is the directional information contained in the acceleration vector because a repositioned errant probe moves in a direction that generally is different from its initial trajectory.

## 4. Repositioning with Directional Information

The straight line defined by $\vec{S}(\eta) = \vec{R}_{j-1}^p + \eta(\vec{R}_j^p - \vec{R}_{j-1}^p)$, $-\infty \leq \eta \leq \infty$, passing through probe $p$'s successive positions $\vec{R}_{j-1}^p$ and $\vec{R}_j^p$ (note $\vec{S}(0) = \vec{R}_{j-1}^p$, $\vec{S}(1) = \vec{R}_j^p$) intersects each of the boundary planes $P_{ik}: X_{ik}$ at $\vec{S}_{ik} = \vec{S}(\eta_{ik}) = \vec{R}_{j-1}^p + \eta_{ik}(\vec{R}_j^p - \vec{R}_{j-1}^p)$. Fig. 1 illustrates this geometry in two dimensions (2D). Planes $P_{11}$, $P_{12}$, $P_{21}$, and $P_{22}$ bound $\Omega$. Point $\vec{R}_{j-1}^p$ lies inside $\Omega$, while $\vec{R}_j^p$ is outside. The acceleration $\vec{a}_{j-1}^p$ has flown probe $p$ outside $\Omega$, so that it must be repositioned somewhere inside the decision space. If the probe's trajectory is constrained to lie along the direction of the acceleration, then its maximum displacement is $d_{max}$, that is, the "distance" between the probe's starting point inside $\Omega$ and the closest boundary plane in the direction of the acceleration as shown in the figure.

The coefficients $\eta_{ik}$ at the points where $\vec{S}(\eta)$ and $P_{ik}$ intersect are determined by the requirement $\hat{e}_i \cdot \vec{S}_{ik} = X_{ik} = \hat{e}_i \cdot \{\vec{R}_{j-1}^p + \eta_{ik}(\vec{R}_j^p - \vec{R}_{j-1}^p)\} = x_i^{p,j-1} + \eta_{ik}(x_i^{p,j} - x_i^{p,j-1})$, so that $\eta_{ik} = \dfrac{X_{ik} - x_i^{p,j-1}}{x_i^{p,j} - x_i^{p,j-1}}$. The *nearest* intersection point in the direction of $p$'s motion is $\vec{S}^* = \vec{S}(\eta^*) = \vec{R}_{j-1}^p + \eta^*(\vec{R}_j^p - \vec{R}_{j-1}^p)$ where $\eta^* = \text{MIN}_{>0}(\eta_{ik})$, $0 \leq \eta^* \leq 1$. Thus, probe $p$'s maximum displacement, which corresponds to placing it on the decision space boundary, is $d_{max} = \|\vec{S}^* - \vec{R}_{j-1}^p\| = \eta^* \|\vec{R}_j^p - \vec{R}_{j-1}^p\|$.

The new repositioning scheme used in this note consequently still utilizes $F_{rep}$, but includes directionality follows:

$$\text{If } \vec{R}_{j-1}^p \in \Omega \text{ and } \vec{R}_j^p \notin \Omega \;\therefore\; \vec{R}_j^p = \vec{R}_{j-1}^p + F_{rep} \, d_{max} \, \hat{a}_{j-1}^p \quad (4),$$

where $\hat{a}_{j-1}^p = \dfrac{\vec{a}_{j-1}^p}{\left\|\vec{a}_{j-1}^p\right\|}$ is the unit vector in the direction of probe $p$'s trajectory between steps $j-1$ and $j$. Because $0 < F_{rep}^{min} \leq F_{rep} \leq 1$, the probe cannot be displaced more than $d_{max}$, thus constraining it to lie inside $\Omega$ or on the boundary; and the unit vector $\hat{a}_{j-1}^p$ insures that $p$ moves only in the direction of its initial acceleration. The value of $F_{rep}$ is set arbitrarily in the same way as before.

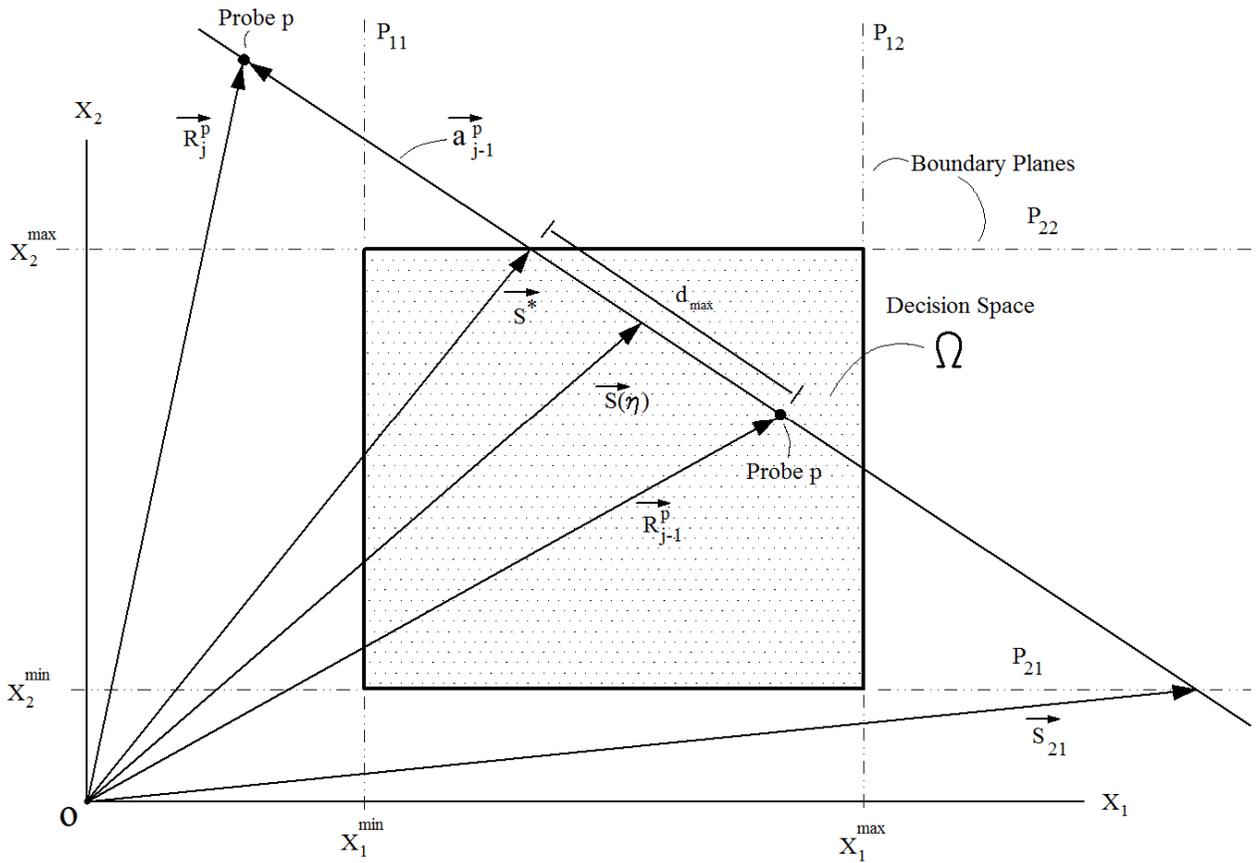

**Fig. 1. Errant probe in a 2D decision space.**

## 5. CFO Implementation

Pseudocode for this implementation is in Fig.2 [Appendix has core routine source code, and a complete electronic listing is available on request, rf2@ieee.org.]. Note that the user inputs only the decision space parameters and the objective function to be maximized.

```
Procedure  CFO [f(x⃗), N_d, Ω]
Internals:  N_t, F_rep^init, ΔF_rep, F_rep^min, (N_p/N_d)_MAX, γ_start, γ_stop, Δγ.

Initialize  f_max^global(x⃗) = very large negative number, say, -10^+4200.

For  N_p/N_d = 2 to (N_p/N_d)_MAX  by  2:
  (a.0)   Total number of probes:  N_p = N_d · (N_p/N_d)

For  γ = γ_start to γ_stop by Δγ:
  (a.1)   Re-initialize data structures for position/
          acceleration vectors & fitness matrix.
  (a.2)   Compute IPD (see [6,11-13]).
  (a.3)   Compute initial fitness matrix, M_0^p, 1 ≤ p ≤ N_p.
  (a.4)   Initialize  F_rep = F_rep^init.

For  j = 0 to N_t  [or earlier termination - see text]:
  (b)     Compute position vectors, R⃗_j^p, 1 ≤ p ≤ N_p [eq.(1)].
  (c)     Retrieve errant probes (1 ≤ p ≤ N_p):
          Selection criteria for methods(c.1)/(c.2)[see text]:
    (c.1) Without directional information:
          If  R⃗_j^p · ê_i < x_i^min  ∴  R⃗_j^p · ê_i = max{x_i^min + F_rep(R⃗_{j-1}^p · ê_i - x_i^min), x_i^min}
          If  R⃗_j^p · ê_i > x_i^max  ∴  R⃗_j^p · ê_i = min{x_i^max - F_rep(x_i^max - R⃗_{j-1}^p · ê_i), x_i^max}
    (c.2) With directional information:
          If  R⃗_{j-1}^p ∈ Ω  and  R⃗_j^p ∉ Ω  ∴  R⃗_j^p = R⃗_{j-1}^p + F_rep d_max â_{j-1}^p
  (d)     Compute fitness matrix for current probe
          distribution,  M_j^p, 1 ≤ p ≤ N_p.
  (e)     Compute accelerations using current probe
          distribution and fitnesses [eq. (2)].
  (f)     Increment F_rep :  F_rep = F_rep + ΔF_rep; If F_rep > 1 ∴ F_rep = F_rep^min.
  (g)     If  j ≥ 20 and j MOD 10 = 0  ∴
          (i)  Shrink Ω around R⃗_best (see [11-13] for details).
          (ii) Retrieve errant probes [procedure Step (c)].
Next j
  (h)     Reset Ω boundaries [values before shrinking].
  (i)     If  f_max(x⃗) ≥ f_max^global(x⃗)  ∴  f_max^global(x⃗) = f_max(x⃗).
Next γ
Next N_p/N_d
```

**Fig. 2. CFO Pseudocode.**

## 6.  Results

Table 1 shows results for CFO with and without directionality in the probe repositioning for a suite of 23 benchmark functions. It compares CFO directly to the Group Search Optimizer (GSO) algorithm [16,17] and indirectly to Particle Swarm (PSO) and Genetic Algorithm (GA) algorithms, which were used in [16] for comparison with GSO. Details of the algorithms, the benchmark suite, and the experimental setup appear there.

The first CFO column shows results with the usual $F_{rep}$ repositioning scheme (eq. (3), no directional information, data reproduced from [13]). The next two columns contain results for directional information using the scheme in eq. (4) applied in two ways: (i) at every step, and (ii) at every other step, respectively. The best fitnesses are highlighted in red, but no highlighting is applied if CFO and the best other algorithm (GSO, PSO, or GA) return essentially the same fitnesses. Note that performance of the other algorithms is described statistically because they are all inherently stochastic, whereas CFO's results are based on a single run because CFO is inherently deterministic. $N_{eval}$ is the total number of CFO function evaluations.

**Table 1.  CFO with/without directional information applied to GSO benchmarks.**

| $F^{(1)}$ | $N_d$ | $F_{max}^{(1)}$ | Avg Best Fitness / Other Algorithm$^{(2)}$ | Central Force Optimization | | | | | |
|---|---|---|---|---|---|---|---|---|---|
| | | | | Reposition without Directional Info$^{(3)}$ | | Reposition with Directional Info$^{(4)}$ | | Reposition with Mixed Directional Info$^{(5)}$ | |
| | | | | Best Fitness | $N_{eval}$ | Best Fitness | $N_{eval}$ | Best Fitness | $N_{eval}$ |
| Unimodal (average of 1000 runs) | | | | Results for a single run because CFO is deterministic. | | | | | |
| $F_1$ | 30 | 0 | -3.6927x10$^{-37}$ / PSO | 0 | 222,960 | 0 | 837,780 | 0 | 616,020 |
| $F_2$ | 30 | 0 | -2.9168x10$^{-24}$ / PSO | 0 | 237,540 | 0 | 764,340 | 0 | 757,920 |
| $F_3$ | 30 | 0 | -1.1979x10$^{-3}$ / PSO | -6.1861x10$^{-5}$ | 397,320 | 0 | 773,400 | -1.5468x10$^{-7}$ | 1,292,820 |
| $F_4$ | 30 | 0 | -0.1078 / GSO | 0 | 484,260 | 0 | 205,980 | 0 | 317,220 |
| $F_5$ | 30 | 0 | -37.3582 / PSO | -4.8623x10$^{-5}$ | 436,680 | -3.1836 | 1,649,580 | -7.8625x10$^{-4}$ | 1,058,700 |
| $F_6$ | 30 | 0 | -1.6000x10$^{-2}$ / GSO | 0 | 176,580 | 0 | 336,420 | 0 | 317,220 |
| $F_7$ | 30 | 0 | -9.9024x10$^{-3}$ / PSO | -1.2919x10$^{-4}$ | 399,960 | -1.7007x10$^{-4}$ | 399,960 | -8.4830x10$^{-4}$ | 397,620 |
| Multimodal, Many Local Max. (avg 1000 runs) | | | | Results for a single run because CFO is deterministic. | | | | | |
| $F_8$ | 30 | 12,569.5 | 12,569.4882 / GSO | 12,569.4865 | 415,500 | 12,569.4551 | 633,420 | 12,569.4376 | 626,220 |
| $F_9$ | 30 | 0 | -0.6509 / GA | 0 | 397,080 | 0 | 745,560 | 0 | 347,880 |
| $F_{10}$ | 30 | 0 | -2.6548x10$^{-5}$ / GSO | 4.7705x10$^{-18}$ | 518,820 | 4.7705x10$^{-18}$ | 627,600 | 4.7705x10$^{-18}$ | 730,020 |
| $F_{11}$ | 30 | 0 | -3.0792x10$^{-2}$ / GSO | -1.7075x10$^{-2}$ | 235,800 | -9.3373x10$^{-2}$ | 185,520 | -2.1875x10$^{-2}$ | 192,360 |
| $F_{12}$ | 30 | 0 | -2.7648x10$^{-11}$ /GSO | -2.1541x10$^{-5}$ | 292,080 | -1.4003x10$^{-3}$ | 729,720 | -1.5237x10$^{-4}$ | 437,220 |
| $F_{13}$ | 30 | 0 | -4.6948x10$^{-5}$ / GSO | -1.8293x10$^{-3}$ | 360,000 | -3.9782x10$^{-2}$ | 945,420 | -8.8244x10$^{-3}$ | 676,020 |
| Multimodal, Few Local Maxima (avg 50 runs) | | | | Results for a single run because CFO is deterministic. | | | | | |
| $F_{14}$ | 2 | -1 | -0.9980 / GSO | -0.9980 | 78,176 | -0.9980 | 83,812 | -0.9980 | 79,716 |
| $F_{15}$ | 4 | -3.075x10$^{-4}$ | -3.7713x10$^{-4}$ / GSO | -5.6967x10$^{-4}$ | 143,152 | -3.4511x10$^{-4}$ | 121,408 | -4.8177x10$^{-4}$ | 426,264 |
| $F_{16}$ | 2 | 1.0316285 | 1.031628 / GSO | 1.03158 | 87,240 | 1.031627 | 88,560 | 1.031628 | 86,504 |
| $F_{17}$ | 2 | -0.398 | -0.3979 / GSO | -0.3979 | 82,096 | -0.3979 | 80,240 | -0.3979 | 94,316 |
| $F_{18}$ | 2 | -3 | -3 / GSO | -3 | 100,996 | -3 | 105,544 | -3 | 100,472 |
| $F_{19}$ | 3 | 3.86 | 3.8628 / GSO | 3.8628 | 160,338 | 3.8628 | 155,010 | 3.8627 | 174,846 |
| $F_{20}$ | 6 | 3.32 | 3.2697 / GSO | 3.3219 | 457,836 | 3.3219 | 439,656 | 3.3219 | 491,076 |
| $F_{21}$ | 4 | 10 | 7.5439 / PSO | 10.1532 | 251,648 | 10.1532 | 243,384 | 10.1532 | 216,864 |
| $F_{22}$ | 4 | 10 | 8.3553 / PSO | 10.4029 | 316,096 | 10.4029 | 303,664 | 10.4029 | 238,344 |
| $F_{23}$ | 4 | 10 | 8.9439 / PSO | 10.5364 | 304,312 | 10.5364 | 326,264 | 10.5364 | 271,816 |

[1] Negative of the functions in [16] are computed by CFO because CFO searches for maxima instead of minima.
[2] Data reproduced from [16].
[3] Data reproduced from [13].
[4] Directional information used on every step.
[5] Directional information used on alternate steps.

The (perhaps) surprising conclusion from these data is that, generally, adding directionality to the original $F_{rep}$ repositioning scheme does not improve convergence, either in terms of speed or accuracy. Figs. 3 and 4 plot convergence speed across the benchmark suite for the two ways in which directional information is included. The charts show the fractional change in the number of function evaluations, $1 - \frac{N'_{eval}}{N_{eval}}$, where $N_{eval}$ is the number of evaluations without directional information and $N'_{eval}$ the number with it. A value of zero corresponds to no change in convergence speed, while positive values reflect improvement and negative ones degradation.

For the group of low dimensionality multimodal functions with few local maxima ($F_{14}$-$F_{23}$), adding directionality at every step (Fig. 3) makes only a slight difference in convergence, with improvement on 6 of the 10 functions. A similar behavior occurs when directional information is applied on alternate steps (Fig. 4), but the variability is greater, and in one case ($F_{15}$) the degradation is substantial.

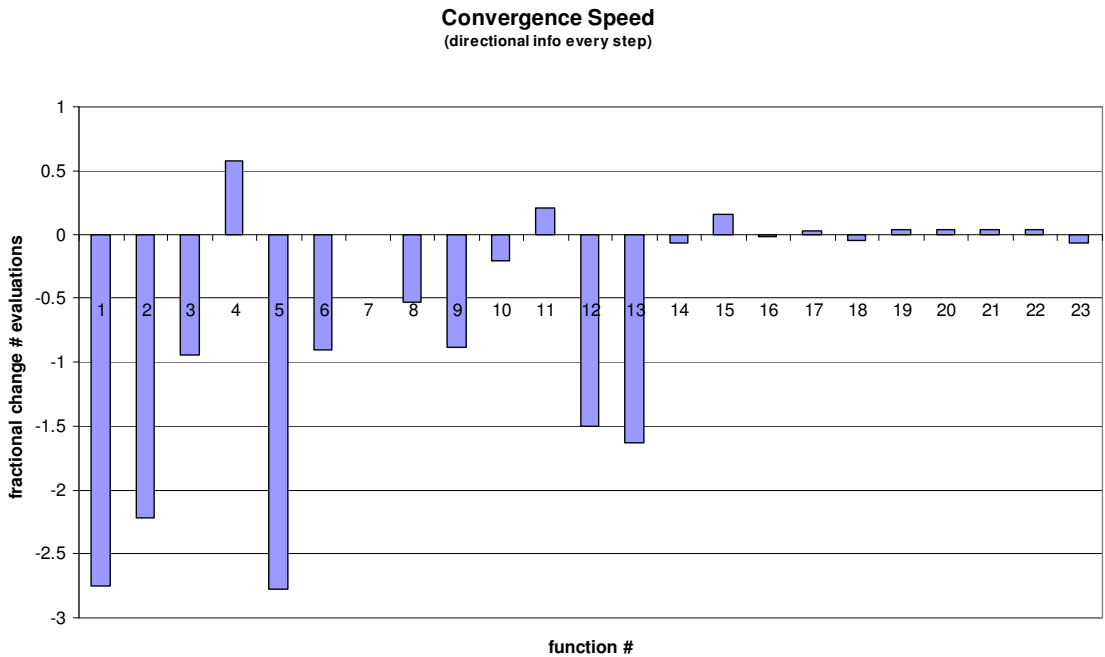

Fig. 3. CFO Convergence speed with directionality on every step (better > 0, worse < 0).

The story is quite different across the two high dimensionality function groups. Adding directionality on the unimodal functions ($F_1$-$F_7$) significantly degrades convergence on five of them with essentially no change on one ($F_7$) and improvement by about a factor of 2 on another ($F_4$). This behavior is evident in both Figs. 3 and 4. A similar trend is seen in the multimodal functions with many local maxima ($F_8$-$F_{13}$). Adding directional information, regardless of the method used, substantially degrades convergence on four functions ($F_8$, $F_{10}$, $F_{12}$, $F_{13}$), while a moderate improvement is seen on one ($F_{11}$). On $F_9$, directionality at every step results in significantly slower convergence, whereas adding it on alternate steps results in a modest improvement.

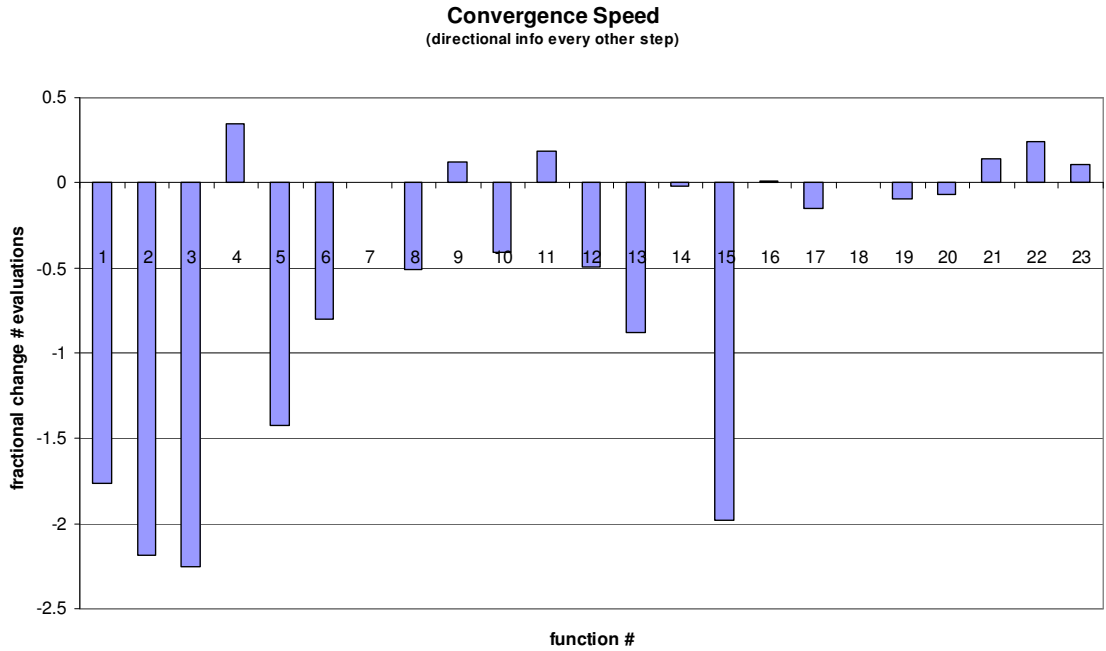

Fig. 4. CFO Convergence speed with directionality on alternate steps (better > 0, worse < 0).

The fitness data in Table 1 show that, for the most part, adding directional information does not materially improve CFO's accuracy. CFO's fitnesses were quite similar in all cases except two. On function $F_3$, directionality at every step results in CFO's locating the known maximum exactly, and including it on alternate steps improves the fitness by more than two orders of magnitude. In this case, adding directional information has improved the accuracy substantially. But in marked contrast, adding directionality on $F_5$ yields very poor results. On this function, whose known maximum is zero, CFO returns a best fitness of $-4.8623 \times 10^{-5}$ with no directional information. Adding directionality on alternate steps results in a fitness of $-7.8625 \times 10^{-4}$, which is worse by a factor of 16. And adding it at every step yields a fitness of $-3.1836$, which is worse yet, by a huge factor of 65,475. This example illustrates that including directionality in some cases can be a *substantial* impediment to CFO's performance. The conflicting results for $F_3$ and $F_5$, which are quite disparate and no doubt a consequence of the functions' unusual topologies, suggests that a *very* cautious approach should be taken to including directionality in CFO.

## 7.    Conclusion

This note compares CFO's performance using the original $F_{rep}$ errant probe retrieval scheme that contains no directional information and using a modified scheme that adds directionality in two ways: (i) at every step and (ii) at every other step. It was speculated that doing so would improve the algorithm's performance, both in terms of convergence speed and accuracy. But the results are somewhat surprising and decidedly contrary to that expectation. For the most part, adding directionality worsens convergence speed. In a very limited number of cases it materially improves speed; but, by far, in most cases convergence is substantially slower. The effect on CFO's accuracy is less pronounced, with the best fitnesses being the same or very similar across the great majority of test functions whether or not directionality is included. In two cases, significant changes are observed, one with a marked increase in accuracy, but the other with a far worse decrease in accuracy.

It appears that there is little if any benefit from adding directionality to CFO's original $F_{rep}$ repositioning scheme, and that doing so likely will slow convergence and, in some cases depending on the objective function's topology, will result in much worse accuracy. Of course, other completely different errant probe repositioning methodologies might benefit from probe directional information, and such different approaches should be developed and investigated. The results reported here hopefully will encourage work in this area.

*30 May 2010*
*Brewster, Massachusetts*

Ver. 2 (Fig. 1 improved for clarity; minor typos corrected.)
*6 June 2010*
*Brewster, Massachusetts*

# Appendix: Source code listing for core CFO routines.

Note: see references [11,13] for complete listings that contain the procedures not included below.

```
'Program 'CFO_04-07-2010(MiniBenchmarks)_VER2.BAS' compiled with
'Power Basic/Windows Compiler 9.04.0122 (www.PowerBasic.com).

'LAST MOD 04-07-2010 ~2116 HRS EDT
'===================================================================================================
#COMPILE EXE
#DIM ALL
%USEMACROS = 1
#INCLUDE "Win32API.inc"
DEFEXT A-Z
'===================================================================================================
'----- MAIN PROGRAM ------
FUNCTION PBMAIN () AS LONG
'   ------ CFO Parameters -----
    LOCAL Nd%, Np%, Nt&
    LOCAL G, DeltaT, Alpha, Beta, Frep AS EXT
    LOCAL PlaceInitialProbes$, InitialAcceleration$, RepositionFactor$
    LOCAL R(), A(), M(), Rbest(), Mbest() AS EXT 'position, acceleration & fitness matrices
    LOCAL FunctionName$  'name of objective function
'   ------------------------- Miscellaneous Setup Parameters ----------------------
    LOCAL N%, i%, YN&, Neval&&, NevalTotal&&, BestNpNd%, NumTrajectories%, Max1DprobesPlotted%, LastStepBestRun&, Pass%
    LOCAL A$, RunCFO$, NECfileError$
    LOCAL BestGamma, BestFitnessThisRun, BestFitnessOverall, StartTime, StopTime AS EXT
    LOCAL BestProbeNum%, BestTimeStep&, BestProbeNumberOverall%, BestTimeStepOverall&, StatusWindowHandle???
'   ------------------- Global Constants --------------------
    REDIM Aij(1 TO 2, 1 TO 25) '(GLOBAL array for Shekel's Foxholes function)
    CALL FillArrayAij
    CALL MathematicalConstants 'NOTE: Calling order is important!!
    CALL AlphabetAndDigits
    CALL SpecialSymbols
    CALL EMconstants
    CALL ConversionFactors        ': CALL ShowConstants 'to verify constants have been set
'   ------------------------- General Setup ---------------------------
    CFOversion$ = "CFO Ver. 04-06-2010"
    RANDOMIZE TIMER  'seed random number generator with program start time
    DESKTOP GET SIZE TO ScreenWidth&, ScreenHeight&  'get screen size (global variables)
    IF DIR$("wgnuplot.exe") = "" THEN
        MSGBOX("WARNING!  'wgnuplot.exe' not found.  Run terminated.") : EXIT FUNCTION
    END IF
'   ---------------------------------------------------- CFO RUN PARAMETERS -----------------------------------------------------
    CALL GetTestFunctionNumber(FunctionName$)' : exit function 'DEBUG
    CALL GetFunctionRunParameters(FunctionName$,Nd%,Np%,Nt&,G,DeltaT,Alpha,Beta,Frep,R(),A(),M(),DiagLength,PlaceInitialProbes$,InitialAcceleration$,RepositionFactor$) 'NOTE: Parameters returned but not used in this version!!
    REDIM R(1 TO Np%, 1 TO Nd%, 0 TO Nt&), A(1 TO Np%, 1 TO Nd%, 0 TO Nt&), M(1 TO Np%, 0 TO Nt&) 'position, acceleration & fitness matrices
    REDIM Rbest(1 TO Np%, 1 TO Nd%, 0 TO Nt&), Mbest(1 TO Np%, 0 TO Nt&) 'overall best position & fitness matrices
'   -------- PLOT 1D and 2D FUNCTIONS ON-SCREEN FOR VISUALIZATION --------
    IF Nd% = 2 AND INSTR(FunctionName$,"PBM_") > 0 THEN
        CALL CheckNECFiles(NECfileError$)
        IF NECfileError$ = "YES" THEN
            EXIT FUNCTION
        ELSE
            MSGBOX("Begin computing plot of function "+FunctionName$+"?  May take a while – be patient...")
        END IF
    END IF
    SELECT CASE Nd%
        CASE 1 : CALL Plot1Dfunction(FunctionName$,R()) : REDIM R(1 TO Np%, 1 TO Nd%, 0 TO Nt&) 'erases coordinate data in R()used to plot function
        CASE 2 : CALL Plot2Dfunction(FunctionName$,R()) : REDIM R(1 TO Np%, 1 TO Nd%, 0 TO Nt&) 'ditto
```

```
        END SELECT

'  -------------------------------------------------- RUN CFO ---------------------------------------------------------

    YN& = MSGBOX("RUN CFO ON FUNCTION " + FunctionName$ + "?"+CHR$(13)+CHR$(13)+"Get some coffee & sit back...",%MB_YESNO,"CONFIRM RUN") : IF YN& = %IDYES THEN RunCFO$ = "YES"

    IF RunCFO$ = "YES" THEN

        StartTime = TIMER

        CALL CFO(FunctionName$,Nd%,Nt&,R(),A(),M(),DiagLength,BestFitnessOverall,BestNpNd%,BestGamma,Neval&&,Rbest(),Mbest(),BestProbeNumberOverall%,BestTimeStepOverall&,LastStepBestRun&,Alpha,Beta)

        StopTime = TIMER

        CALL PlotResults(FunctionName$,Nd%,BestFitnessOverall,BestNpNd%,BestGamma,Neval&&,Rbest(),Mbest(),BestProbeNumberOverall%,BestTimeStepOverall&,LastStepBestRun&,Alpha,Beta)

        MSGBOX(FunctionName$+CHR$(13)+"Total Function Evaluations = "+STR$(Neval&&)+CHR$(13)+"Runtime = "+STR$(ROUND((Stoptime-StartTime)/3600##,2))+" hrs")
    END IF

ExitPBMAIN:

END FUNCTION 'PBMAIN()

'============================================================= CFO SUBROUTINE =========================================================
SUB CFO(FunctionName$,Nd%,Nt&,R(),A(),M(),DiagLength,BestFitnessOverall,BestNpNd%,BestGamma,Neval&&,Rbest(),Mbest(),BestProbeNumberOverall%,BestTimeStepOverall&,LastStepBestRun&,Alpha,Beta)

LOCAL p%, i%, j& 'Standard Indices: Probe #, Coordinate #, Time Step #

LOCAL Np% 'Number of Probes

LOCAL MaxProbesPerDimension% 'Maximum # probes per dimension (depends on Nd%)

LOCAL k%, L% 'Dummy summation indices

LOCAL NumProbesPerDimension%, GammaNumber%, NumGammas% 'Probes/dimension on each probe line; gamm point #; # gamma points

LOCAL SumSQ, Denom, Numerator, Gamma, Frep, DeltaFrep AS EXT

LOCAL BestProbeNumber%, BestTimeStep&, LastStep&, BestProbeNumberThisRun%, BestTimeStepThisRun&

LOCAL BestFitness, BestFitnessThisRun, eta AS EXT

LOCAL FitnessSaturation$

LOCAL ReposInterval%

'---------------- Initial Parameter Values ------------------

ReposInterval% = 2

BestFitnessOverall = -1E4200 'very large NEGATIVE number...

Neval&& = 0 'function evaluation counter

'----------------------- CFO Parameters --------------------

Alpha = 1## : Beta = 1## 'CFO EXPONENTS Alpha & Beta

Nt& = 1000 'set to a large value expecting early termination

IF FunctionName$ = "F7" THEN Nt& = 100 'to reduce runtime because this function contains random noise

LastStep& = Nt&

DeltaFrep = 0.1## '0.1##

NumGammas% = 11

SELECT CASE Nd% 'set Np%/Nd% based on Nd% to avoid excessive runtimes

    CASE   1 TO   6 : MaxProbesPerDimension% = 14
    CASE   7 TO  10 : MaxProbesPerDimension% = 12
    CASE  11 TO  15 : MaxProbesPerDimension% = 10
    CASE  16 TO  20 : MaxProbesPerDimension% =  8
    CASE  21 TO  30 : MaxProbesPerDimension% =  6
    CASE ELSE       : MaxProbesPerDimension% =  4
END SELECT

'   ------------------------- Np/Nd LOOP --------------------

FOR NumProbesPerDimension% = 2 TO MaxProbesPerDimension% STEP 2

Np% = NumProbesPerDimension%*Nd%

'   ------------------------- GAMMA LOOP --------------------

FOR GammaNumber% = 1 TO NumGammas%

Gamma = (GammaNumber%-1)/(NumGammas%-1)

REDIM R(1 TO Np%, 1 TO Nd%, 0 TO Nt&), A(1 TO Np%, 1 TO Nd%, 0 TO Nt&), M(1 TO Np%, 0 TO Nt&) 're-initializes Position Vector/Acceleration/Fitness matrices to zero

'STEP (A1) ------------- Compute Initial Probe Distribution (Step 0)----------------

    CALL IPD(Np%,Nd%,Nt&,R(),Gamma) 'Probe Line IPD intersecting on diagonal at a point determined by Gamma

'STEP (A2) ------------- Compute Initial Fitness Matrix (Step 0) ------------------------

    FOR p% = 1 TO Np% : M(p%,0) = ObjectiveFunction(R(),Nd%,p%,0,FunctionName$) : INCR Neval&& : NEXT p%

'STEP (A3) ------------- Set Initial Probe Accelerations to ZERO (Step 0)----------------
```

```
    FOR p% = 1 TO Np% : FOR i% = 1 TO Nd% : A(p%,i%,0) = 0## : NEXT i% : NEXT p%

'STEP (A4) ------------- Initialize Frep ----------------

    Frep = 0.5##

' ===================================== LOOP ON TIME STEPS STARTING AT STEP #1 =========================================

    BestFitnessThisRun = -1E4200 'very large NEGATIVE number

    FOR j& = 1 TO Nt&

'STEP (B) ----------- Compute Probe Position Vectors for this Time Step --------

        FOR p% = 1 TO Np% : FOR i% = 1 TO Nd% : R(p%,i%,j&) = R(p%,i%,j&-1) + A(p%,i%,j&-1) : NEXT i% : NEXT p% 'note: factor of 1/2 combined with G=2 to produce coefficient of one

'STEP (C) ----------- Retrieve Errant Probes ---------------

        IF j& MOD ReposInterval% = 0 THEN 'use directional scheme 1/3 of the time
            CALL RetrieveErrantprobes2(Np%,Nd%,j&,R(),A(),Frep) 'added 04-01-10
        ELSE
            CALL RetrieveErrantProbes(Np%,Nd%,j&,R(),Frep)
        END IF

'STEP (D) ----------- Compute Fitness Matrix for Current Probe Distribution ---------

        FOR p% = 1 TO Np% : M(p%,j&) = ObjectiveFunction(R(),Nd%,p%,j&,FunctionName$) : INCR Neval&& : NEXT p%

'STEP (E) ----------- Compute Accelerations Based on Current Probe Distribution & Fitnesses ---------------

        FOR p% = 1 TO Np%

            FOR i% = 1 TO Nd%

                A(p%,i%,j&) = 0

                FOR k% = 1 TO Np%

                    IF k% <> p% THEN

                        SumSQ = 0## : FOR L% = 1 TO Nd% : SumSQ = SumSQ + (R(k%,L%,j&)-R(p%,L%,j&))^2 : NEXT L% 'dummy index

                        IF SumSQ <> 0## THEN 'to avoid zero denominator (added 03-20-10)

                            Denom = SQR(SumSQ) : Numerator = UnitStep((M(k%,j&)-M(p%,j&)))*(M(k%,j&)-M(p%,j&))

                            A(p%,i%,j&) = A(p%,i%,j&) + (R(k%,i%,j&)-R(p%,i%,j&))*Numerator^Alpha/Denom^Beta 'ORIGINAL VERSION WITH VARIABLE Alpha & Beta

                        END IF 'added 03-20-10

                    END IF

                NEXT k% 'dummy index

            NEXT i% 'coord (dimension) #

        NEXT p% 'probe #

' --------- Get Best Fitness & Corresponding Probe # and Time Step ---------

    CALL GetBestFitness(M(),Np%,j&,BestFitness,BestProbeNumber%,BestTimeStep&)

    IF BestFitness >= BestFitnessThisRun THEN

        BestFitnessThisRun = BestFitness : BestProbeNumberThisRun% = BestProbeNumber% : BestTimeStepThisRun& = BestTimeStep&

    END IF

' ----- Increment Frep -----

    Frep = Frep + DeltaFrep

    IF Frep > 1## THEN Frep = 0.05## 'keep Frep in range [0.05,1]

' ---------- Starting at Step #20 Shrink Decision Space Around Best Probe Every 20th Step -----------

    IF j& MOD 20 = 0 AND j& >= 20 THEN

        FOR i% = 1 TO Nd% : XiMin(i%) = XiMin(i%)+(R(BestProbeNumber%,i%,BestTimeStep&)-XiMin(i%))/2## : XiMax(i%) = XiMax(i%)-(XiMax(i%)-R(BestProbeNumber%,i%,BestTimeStep&))/2## : NEXT i% 'shrink DS by 0.5

        IF j& MOD ReposInterval% = 0 THEN
            CALL RetrieveErrantprobes2(Np%,Nd%,j&,R(),A(),Frep) 'added 04-01-10
        ELSE
            CALL RetrieveErrantProbes(Np%,Nd%,j&,R(),Frep) 'TO RETRIEVE PROBES LYING OUTSIDE SHRUNKEN DS 'ADDED 02-07-2010
        END IF

    END IF

' ---------- If SlopeRatio Changes Abruptly, Shrink Decision Space Around Best Probe ------ 'SOMETIMES SEEMS TO IMPROVE PERFORMANCE, BUT NOT WITH VERY ERRATIC Fitness Evolution PLOTS...

'    eta = 0.9##

'    IF SlopeRatio(M(),Np%,j&) >= 3## THEN

'        FOR i% = 1 TO Nd% : XiMin(i%) = XiMin(i%)+eta*(R(BestProbeNumber%,i%,BestTimeStep&)-XiMin(i%)) : XiMax(i%) = XiMax(i%)-eta*(XiMax(i%)-R(BestProbeNumber%,i%,BestTimeStep&)): NEXT i%

'        if j& mod ReposInterval% = 0 then
'            CALL RetrieveErrantprobes2(Np%,Nd%,j&,R(),A(),Frep) 'added 04-01-10
'        else
'            CALL RetrieveErrantProbes(Np%,Nd%,j&,R(),Frep) 'TO RETRIEVE PROBES LYING OUTSIDE SHRUNKEN DS 'ADDED 02-07-2010
'        end if

'    END IF

'STEP (F) --------------- Check for Early Run Termination ----------------

    IF HasFITNESSsaturated$(25,j&,Np%,Nd%,M(),R(),DiagLength) = "YES" THEN

        LastStep& = j& : EXIT FOR 'exit TIME STEP Loop
```

```
            END IF
        NEXT j& 'END TIME STEP LOOP
'------------------------------------------------- Best Overall Fitness & Corresponding Run Parameters ----------------------------------
IF BestFitnessThisRun >= BestFitnessOverall THEN
        BestFitnessOverall = BestFitnessThisRun : BestProbeNumberOverall% = BestProbeNumberThisRun% : BestTimeStepOverall& = BestTimeStepThisRun&
        BestNpNd% = NumProbesPerDimension% : BestGamma = Gamma : LastStepBestRun& = LastStep&
        CALL CopyBestMatrices(Np%,Nd%,Nt&,R(),M(),Rbest(),Mbest())
END IF
'STEP (G) ----- Reset Decision Space Boundaries to Initial Values -----
CALL ResetDecisionSpaceBoundaries(Nd%)
NEXT GammaNumber% 'END GAMMA LOOP
NEXT NumProbesPerDimension% 'END Np/Nd LOOP
END SUB 'CFO()
'==============================================================================================================================
SUB IPD(Np%,Nd%,Nt&,R(),Gamma) 'Initial Probe Distribution (IPD) on "Probe Lines" Parallel to Coordinate Axes
LOCAL DeltaXi, DelX1, DelX2, Di AS EXT
LOCAL NumProbesPerDimension%, p%, i%, k%, NumX1points%, NumX2points%, x1pointNum%, x2pointNum%
            IF Nd% > 1 THEN
                NumProbesPerDimension% = Np%\Nd% 'even #
            ELSE
                NumProbesPerDimension% = Np%
            END IF
            FOR i% = 1 TO Nd%
                FOR p% = 1 TO Np%
                    R(p%,i%,0) = XiMin(i%) + Gamma*(XiMax(i%)-XiMin(i%))
                NEXT Np%
            NEXT i%
            FOR i% = 1 TO Nd% 'place probes probe line-by-probe line (i% is dimension [coordinate] number)
                DeltaXi = (XiMax(i%)-XiMin(i%))/(NumProbesPerDimension%-1)
                FOR k% = 1 TO NumProbesPerDimension%
                    p% = k% + NumProbesPerDimension%*(i%-1) 'probe #
                    R(p%,i%,0) = XiMin(i%) + (k%-1)*DeltaXi
                NEXT k%
            NEXT i%
END SUB 'IPD()
'-------------
SUB GetBestFitness(M(),Np%,StepNumber&,BestFitness,BestProbeNumber%,BestTimeStep&)
LOCAL p%, jj&, A$
    BestFitness = -1E4200 'very large negative number
    FOR jj& = 0 TO StepNumber&
        FOR p% = 1 TO Np%
            IF M(p%,jj&) >= BestFitness THEN
                BestFitness = M(p%,jj&) : BestProbeNumber% = p% : BestTimeStep& = jj&
            END IF
        NEXT p%
    NEXT jj&
END SUB 'GetBestFitness()
'------------------------
FUNCTION HasFITNESSsaturated$(NavgSteps&,j&,Np%,Nd%,M(),R(),DiagLength)
LOCAL A$, B$
LOCAL k&, p%
LOCAL BestFitness, SumOfBestFitnesses, BestFitnessStepJ, FitnessSatTOL AS EXT
    A$ = "NO" : B$ = "j="+STR$(j&)+CHR$(13)
    FitnessSatTOL = 0.000001## 'tolerance for FITNESS saturation
    IF j& < NavgSteps& + 10 THEN GOTO ExitHasFITNESSsaturated 'execute at least 10 steps after averaging interval before performing this check
    SumOfBestFitnesses = 0##
    FOR k& = j&-NavgSteps&+1 TO j& 'GET BEST FITNESSES STEP-BY-STEP FOR NavgSteps& INCLUDING THIS STEP j& AND COMPUTE AVERAGE VALUE.
```

```
'       BestFitness = M(k&,1) 'ORIG CODE 03-23-2010: THIS IS A MISTAKE!

        BestFitness = -1E4200 'THIS LINE CORRECTED 03-23-2010 PER DISCUSSION WITH ROB GREEN.
                              'INITIALIZE BEST FITNESS AT k&-th TIME STEP TO AN EXTREMELY LARGE NEGATIVE NUMBER.

        FOR p% = 1 TO Np% 'PROBE-BY-PROBE GET MAXIMUM FITNESS

            IF M(p%,k&) >= BestFitness THEN BestFitness = M(p%,k&)

        NEXT p%

        IF k& = j& THEN BestFitnessStepJ = BestFitness  'IF AT THE END OF AVERAGING INTERVAL, SAVE BEST FITNESS FOR CURRENT TIME STEP j&

        SumOfBestFitnesses = SumOfBestFitnesses + BestFitness

    NEXT k&

    IF ABS(SumOfBestFitnesses/NavgSteps&-BestFitnessStepJ) =< FitnessSatTOL THEN A$ = "YES" 'saturation if (avg value - last value) are within TOL

ExitHasFITNESSsaturated:

    HasFITNESSsaturated$ = A$

END FUNCTION 'HasFITNESSsaturated$()

'---------------------------------

SUB RetrieveErrantprobes2(Np%,Nd%,j&,R(),A(),Frep) 'added 04-01-10

LOCAL A$, ProbeInsideDSstepJ$, ProbeInsideDSstepJminus1$

LOCAL p%, i%, k%

LOCAL Xik, dMax, EtaIK(), EtaStar, SumSQ, MagRjRj1, MagAj1, Numerator, Denom AS EXT

    FOR p% = 1 TO Np% 'check every probe, probe-by-probe
'       ---------------------- Determine Probe Locations at Steps j& and j&-1 ------------------------------

        ProbeInsideDSstepJ$ = "YES" 'presume probe is inside DS at step j&

        FOR i% = 1 TO Nd% 'check to see if probe p lies outside DS (any coordinate beyond corresponding boundary coordinate)

            IF (R(p%,i%,j&) > XiMax(i%) OR R(p%,i%,j&) < XiMin(i%)) THEN 'probe lies outside DS

                ProbeInsideDSstepJ$ = "NO" : EXIT FOR 'need only one coordinate outside DS

            END IF

        NEXT i%

        ProbeInsideDSstepJminus1$ = "YES" 'presume probe is inside DS at step j&-1

        FOR i% = 1 TO Nd% 'check to see if probe p lies outside DS (any coordinate beyond corresponding boundary coordinate)

            IF (R(p%,i%,j&-1) > XiMax(i%) OR R(p%,i%,j&-1) < XiMin(i%)) THEN 'probe lies outside DS

                ProbeInsideDSstepJminus1$ = "NO" : EXIT FOR 'need only one coordinate outside DS

            END IF

        NEXT i%
'       ------------------------------- If Probe is Outside at Both Steps, Use Old Scheme to Reposition --------------------------------------

        IF ProbeInsideDSstepJ$ = "NO" AND ProbeInsideDSstepJminus1$ = "NO" THEN 'probe p% is outside DS at both time steps => use old Frep scheme to reposition

            FOR i% = 1 TO Nd%

                IF R(p%,i%,j&-1) < XiMin(i%) THEN R(p%,i%,j&-1) = MAX(XiMin(i%) + Frep*(R(p%,i%,j&-2)-XiMin(i%)),XiMin(i%))
                IF R(p%,i%,j&-1) > XiMax(i%) THEN R(p%,i%,j&-1) = MIN(XiMax(i%) - Frep*(XiMax(i%)-R(p%,i%,j&-2)),XiMax(i%))

                IF R(p%,i%,j&)   < XiMin(i%) THEN R(p%,i%,j&)   = MAX(XiMin(i%) + Frep*(R(p%,i%,j&-1)-XiMin(i%)),XiMin(i%))
                IF R(p%,i%,j&)   > XiMax(i%) THEN R(p%,i%,j&)   = MIN(XiMax(i%) - Frep*(XiMax(i%)-R(p%,i%,j&-1)),XiMax(i%))

            NEXT i%

        END IF 'ProbeInsideDSstepJ$ = "NO" AND ProbeInsideDSstepJminus1$ = "NO"
'       -------------- If Probe is Outside at Step j& but Inside at Step j&-1 Then Use Reposition Using Directional Information --------------

        IF ProbeInsideDSstepJ$ = "NO" AND ProbeInsideDSstepJminus1$ = "YES" THEN 'probe p% is outside DS at step j& and inside at step j&-1 => use scheme that preserves directional information

            REDIM EtaIK(1 TO Nd%, 1 TO 2) 'Eta(i%,k%)

            FOR i% = 1 TO Nd% 'compute array of Eta values

                FOR k% = 1 TO 2

                    SELECT CASE k%
                        CASE 1 : Xik = XiMin(i%)
                        CASE 2 : Xik = XiMax(i%)
                    END SELECT

                    Numerator = Xik-R(p%,i%,j&-1) : Denom = R(p%,i%,j&)-R(p%,i%,j&-1)

                    IF ABS(Denom) =< 1E-10 THEN

                        EtaIK(i%,k%) = 0## 'DO NOT REPOSITION

                    ELSE

                        EtaIK(i%,k%) = Numerator/Denom

                    END IF

                NEXT k%

            NEXT i%
```

```
            EtaStar = 1E4200 'very large POSITIVE number

            FOR i% = 1 TO Nd% 'get min Eta value >= 0

                FOR k% = 1 TO 2

                    IF EtaIK(i%,k%) =< EtaStar AND EtaIK(i%,k%) >= 0## THEN EtaStar = EtaIK(i%,k%)

                NEXT k%

            NEXT i%

IF EtaStar < 0## OR EtaStar > 1## THEN MSGBOX("WARNING! EtaStar="+STR$(EtaStar))

            SumSQ = 0## : FOR i% = 1 TO Nd% : SumSQ = SumSQ + (R(p%,i%,j&)- R(p%,i%,j&-1))^2 : NEXT i% : MagRjRj1 = SQR(SumSQ) 'magnitude of
[(Rp at step j) MINUS (Rp at step j-1)]

            dMax = EtaStar*MagRjRj1 'distance to nearest boundary plane from position of probe p% at step j&-1

            SumSQ = 0## : FOR i% = 1 TO Nd% : SumSQ = SumSQ + A(p%,i%,j&-1)^2 : NEXT i% : MagAj1 = SQR(SumSQ) 'magnitude of acceleration at
step j-1

            FOR i% = 1 TO Nd% 'reposition probe p usimng acceleration directional information

                R(p%,i%,j&) = R(p%,i%,j&-1) + Frep*dMax*A(p%,i%,j&-1)/MagAj1 'unit vector in direction of acceleration preserves acceleration
directional information

IF dMax > DiagLength THEN MSGBOX("WARNING! dMax="+STR$(dMax)+"  Diag="+STR$(DiagLength)+"  MAG Aj1="+STR$(MagAj1)+"  Frep="+STR$(Frep)+"
EtaStar="+STR$(EtaStar))

            NEXT i%

        END IF 'ProbeInsideDSstepJ$ = "NO" AND ProbeInsideDSstepJminus1$ = "YES"

    NEXT p% 'process next probe

END SUB 'Retrieveerrantprobes2()

'-------------------------=---

SUB RetrieveErrantProbes(Np%,Nd%,j&,R(),Frep) 'original version, does not include acceleration vector directional information

LOCAL p%, i%

    FOR p% = 1 TO Np%

        FOR i% = 1 TO Nd%

            IF R(p%,i%,j&) < XiMin(i%) THEN R(p%,i%,j&) = MAX(XiMin(i%) + Frep*(R(p%,i%,j&-1)-XiMin(i%)),XiMin(i%)) 'CHANGED 02-07-10

            IF R(p%,i%,j&) > XiMax(i%) THEN R(p%,i%,j&) = MIN(XiMax(i%) - Frep*(XiMax(i%)-R(p%,i%,j&-1)),XiMax(i%))

        NEXT i%

    NEXT p%

END SUB 'RetrieveErrantProbes()

'-----------------------------

SUB ResetDecisionSpaceBoundaries(Nd%)

    LOCAL i%

    FOR i% = 1 TO Nd% : XiMin(i%) = StartingXiMin(i%) : XiMax(i%) = StartingXiMax(i%) : NEXT i%

END SUB 'ResetDecisionSpaceBoundaries()

'-------------------------------------
```

<div style="text-align:center">END OF PARTIAL SOURCE CODE LISTING</div>